\documentclass[12pt]{article}

\input epsf.sty
\usepackage[centertags]{amsmath}
\usepackage{amsfonts}
\usepackage{amssymb}
\usepackage{amsthm}
\usepackage{newlfont}
\usepackage{epsfig}
\usepackage{amscd}
\newtheorem{thm}{Theorem}

\newcommand{\tH}{\theta}

\newcommand{\Di}{\Delta_m}
\newcommand{\Dj}{\Delta_n} 
\newcommand{\Dim}{\Delta_{-m}}
\newcommand{\Djm}{\Delta_{-n}}
\newcommand{\N}{\hat{N}}

\newcommand{\R}{\vec{R}}
\newcommand{\nn}{\hat{n}}
\newcommand{\rr}{\vec{r}}

\newcommand{\G}{\Gamma}

\def\mref#1{(\ref{#1})}
\def\eqref#1{(\ref{#1})}

\begin{document}
\title{\bf The self-adjoint 5-point and 7-point  difference  operators, 
the associated Dirichlet problems, 
 Darboux transformations and Lelieuvre formulas
}
\author{
M. Nieszporski\thanks{
Katedra Metod Matematycznych Fizyki,
Uniwersytet Warszawski 
ul. Ho\.za 74, 00-682 Warszawa, Poland
 e-mail: maciejun@fuw.edu.pl,
tel: +48 22 621 77 57, 
fax: +48 22 622 45 08
}
\thanks{Instytut Fizyki Teoretycznej,
Uniwersytet w Bia{\l}ymstoku,
ul. Lipowa 41, 15-424 Bia{\l}ystok, Poland}~~and ~ 
P. M. Santini\thanks{Dipartimento di Fisica, Universit\`a di Roma ``La Sapienza'' and 
Istituto Nazionale di Fisica Nucleare, Sezione di Roma, 
Piazz.le Aldo Moro 2, I--00185 Roma, Italy
e-mail: paolo.santini@roma1.infn.it}
}

\maketitle

\date

\begin{abstract}
We present some basic properties of two distinguished discretizations 
of elliptic operators: the self-adjoint 5-point 
and 7-point schemes on a two dimensional  lattice. We first show that they allow to solve
Dirichlet boundary value problems; then we present their Darboux transformations. 
Finally we construct their Lelieuvre formulas and 
we show that, at the level of the normal vector and 
in full analogy with their continuous counterparts, the self-adjoint    
5-point scheme characterizes a two dimensional 
quadrilateral lattice (a lattice whose elementary quadrilaterals 
are planar), while the self-adjoint 7-point scheme characterizes  
a generic 2D lattice.
\end{abstract}

\section{Introduction}

In the last two decades of the XIX century and in the beginning of the XX century
many great mathematicians (Bianchi, Darboux and others) developed a 
differential geometry studying transformations of certain geometric structures 
and proved theorems of permutability of such transformations,  obtaining 
in turn nonlinear superposition
principles for the nonlinear differential equations characterizing the 
above geometries. These results can be viewed as the pre-history  
\cite{H,Prus} of the modern theory 
of integrable nonlinear systems, which is based on the existence of linear 
differential operators possessing symmetry transformations of Darboux type (the Darboux 
Transformations (DTs)) and nonlinear isospectral symmetries (the 
integrable nonlinear systems).

After more than one century, in studying discrete integrable systems (which 
are, in a sense, richer and more fundamental than their continuous 
counterparts and, for these reasons, worth studying), one often makes use of 
the mutual interplay between geometry and the theory of integrable systems 
also in the discrete case \cite{Bob}.

The main goal of this paper is to present two examples of such an 
interplay; we start with the self-adjoint 7-point operator 
\begin{equation}
\label{L7}
\begin{array}{l}
{\mathcal L}_{7}:=
a_{m,n}T_m+a_{m-1,n}T^{-1}_m+b_{m,n}T_n+ 
b_{m,n-1}T^{-1}_n+ 
s_{m+1,n}T_m T^{-1}_n+ \\
s_{m,n+1}T^{-1}_mT_n-f_{m,n}
\end{array}
\end{equation}
and with the self-adjoint 5-point operator
\begin{equation}
\label{L5}
{\mathcal L}_{5}:=a_{m,n}T_m+a_{m-1,n}T^{-1}_m+b_{m,n}T_n+b_{m,n-1}T^{-1}_n-
f_{m,n},
\end{equation}
for which the existence of Darboux transformations has been recently 
established \cite{NDS}, and we obtain their geometric interpretation  
through the construction of their Lelieuvre formulas. It is worth 
mentioning that, very 
often, one follows the opposite direction: a ``geometric insight'' allows 
one to construct an integrable system. 

In the above equations 
$T_m$ and $T_n$ are the translation operators with respect to the 
discrete variables $(m,n)\in{\mathbb Z}^2$
\begin{displaymath}
T_mf_{m,n}=f_{m+1,n},~~~T_nf_{m,n}=f_{m,n+1}
\end{displaymath}
and $f_{m,n}=f(m,n)$ is a function of $(m,n)$. 

The classical Lelieuvre formulas \cite{Lelieuvre} 
\begin{equation}
\label{Lelieuvre}
\begin{array}{ccc}
\vec{R},_u= \vec{N},_u \times \vec{N} ,&\qquad &
\vec{R},_v= \vec{N} \times \vec{N},_v.
\end{array}
\end{equation}
allow one to construct a two-dimensional surface $\R(u,v)$
in $E^3$ from its normal (not necessarily the unit one) image $\vec{N}(u,v)$. 
The co-ordinate net $(u,v)$ of the surface obtained in this way is the asymptotic one and 
the normal field satisfies the Moutard equation \cite{Moutard}
\begin{equation}
\vec{N},_{uv}=F(u,v) \vec{N}.
\end{equation}
The Moutard equation is covariant under the Moutard transformation. A  
second Darboux transformation appeared in literature  \cite{Moutard} 
(the first one
was the transformation of Ribaucour); it gave Guichard  \cite{Guichard} the possibility to describe
Weingarten rectilinear congruences in a very elegant way and, in turn, it allowed
Bianchi and other Geometers to construct many systems of nonlinear differential equations
which now are called soliton systems.

Another example of Lelieuvre type formula was obtained by Bianchi
\cite{Bianchi}
\begin{equation}
\begin{array}{ccc}
\vec{R},_x= \vec{N},_y \times \vec{N} ,&\qquad &
\vec{R},_y= \vec{N} \times \vec{N},_x.
\end{array}
\end{equation}
Now the normal field satisfies the 2D Schr\"odinger equation
\begin{equation}
 \vec{N},_{xx}+ \vec{N},_{yy}=F \vec{N},
\end{equation}
which is also covariant under a  Darboux transformation, and 
the coordinate net $(x,y)$ is, in this case, the isothermally-conjugate one.

An extension of the Lelieuvre formulas 
to an arbitrary co-ordinate system (or, better to say, to a co-ordinate free 
language)
and to hyper-surfaces in the equi-affine space of arbitrary dimension has 
been obtained in \cite{LiNo,Li}.
Another extension of the Lelieuvre formulas can be found 
in \cite{Schief}.
A discretization of the Lelieuvre formulas \mref{Lelieuvre} was first given 
in the paper
\cite{Konopelchenko} and a discretization of the notion of Weingarten 
congruences 
was proposed in \cite{Nieszpor}. 

The main result of this paper consists in the construction 
of the Lelieuvre formulas for a 2D quadrilateral lattice (a lattice whose elementary 
quadrilaterals are planar) \cite{Sauer}, \cite{Doliwa} and for an   
essentially arbitrary 2D lattice in $eA^3$ space. This result 
allows one to establish that the operators (\ref{L5}) and 
(\ref{L7}) characterize, at the level of the normal vectors, respectively, 
the above 2D lattices. 

For our purposes it is not necessary to deal with the Euclidean space, but it 
is enough
to enrich the affine space with the volume form $Vol$ (by $Vol^*$ we denote the dual
form of $Vol$); i.e.,  it is enough to deal with the equi-affine space $eA^3$. 
This enables one to construct the cross product 
from an ordered pair of linearly independent vector fields (say $(\vec{a},\vec{b})$); 
i.e., to construct the element $\hat{N} \in T^*eA^3$ such that
$<\hat{N}|\vec{a}>=0=<\hat{N}|\vec{b}>$ and $<\hat{N}|\vec{c}>=Vol \{\vec{a};\vec{b};\vec{c} \}$
for every $\vec{c} \in TeA^3$.

The second goal of this paper consists of the illustration of some of 
the basic criteria for constructing the proper discretizations of 
partial differential operators. Again we use, as illustrative examples,  
the operators (\ref{L7}) and (\ref{L5}).  

The paper is organized as follows.
In section \ref{Criteria} we present some of the  
basic criteria which we use as a guide for constructing  
the proper discretizations of partial differential operators. These 
criteria are systematically applied, in the remaining sections, to the 
illustrative examples given by the operators ${\mathcal L}_{5}$ and ${\mathcal L}_{7}$.
 In section \ref{BV} we show that the operators ${\mathcal L}_{5}$ and 
${\mathcal L}_{7}$ preserve the elliptic character of their differential 
counterpart, being applicable to solve the Dirichlet problem on a 
2D lattice. In section \ref{SAD} we show that the operators 
${\mathcal L}_{5}$ and ${\mathcal L}_{7}$ possess, like their differential 
counterparts, DTs. In sections \ref{CL}, \ref{GL} and \ref{geom} we 
derive the Lelieuvre formulas for, respectively, the continuous counterpart 
of ${\mathcal L}_{7}$, for ${\mathcal L}_{7}$, for the continuous counterpart 
of ${\mathcal L}_{5}$ and for ${\mathcal L}_{5}$, verifying that the geometric 
meaning of the operators ${\mathcal L}_{5}$ and ${\mathcal L}_{7}$ is 
the proper discretization of the geometric meaning of their differential 
counterparts.

We conclude this introductory section with some general remarks on the 
operators ${\mathcal L}_{5}$ and ${\mathcal L}_{7}$. 
The operator ${\mathcal L}_{7}$ can be 
interpreted as the most general self-adjoint 
operator on the star of a regular triangular lattice \cite{Novikov,Novikov2}; it possesses a class 
of Laplace transformations \cite{Novikov,Novikov2} and   
plays a relevant role in a recently developed discrete complex 
function theory \cite{Novikov3}. Its natural continuous limit \cite{NDS}:
\begin{equation}
\label{selffc}
A \partial^2_x+B\partial^2_y+2S\partial_x\partial_y +
(A,_x+S,_y)\partial_x+(B,_y+S,_x)\partial_y-F
\end{equation}
is the most general, second order, linear, self-adjoint operator of the second order. 
The operator ${\mathcal L}_{5}$ is instead 
the most general self-adjoint operator on the star of a square lattice 
\cite{NDS} and its natural continuous limit 
is the following self-adjoint elliptic (if $AB>0$) operator
\begin{equation}
\label{selfc}
A\partial^2_x+A,_x\partial_x+B\partial^2_y+B,_y\partial_y-F.
\end{equation}
It is interesting to remark that the 
following distinguished gauge equivalent form of the operator 
${\mathcal L}_{5}$ \cite{NDS}:
\begin{equation}
\label{schr5}
{\mathcal L}_{SchInt}:=\frac{\G_{m,n}}{\G_{m+1,n}} T_m+
\frac{\G_{m-1,n}}{\G_{m,n}}T^{-1}_m+
\frac{\G_{m,n}}{\G_{m,n+1}} T_n+\frac{\G_{m,n-1}}{\G_{m,n}}T^{-1}_n-q_{m,n},
\end{equation}
admits DTs and reduces, in the continuous limit, to the celebrated 
Schr\"odinger operator in the plane  
\begin{equation}
\label{schr}
\partial^2_x+\partial^2_y-Q.
\end{equation}
Therefore it can be considered as a distinguished integrable discretization 
of the Schr\"odinger operator \cite{NDS}. 

\begin{center}
\mbox{ \fbox{\epsfysize=4cm \epsffile{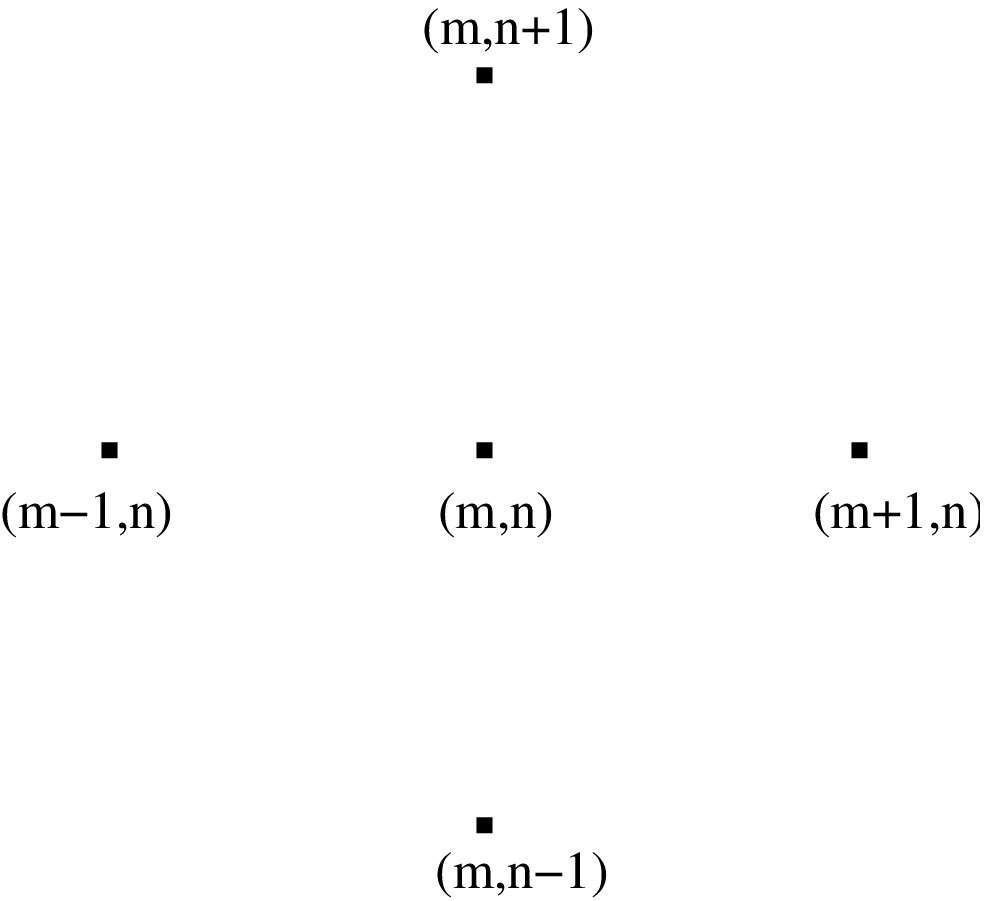}}~~~~~~~~~~~ 
\fbox{\epsfysize=4cm \epsffile{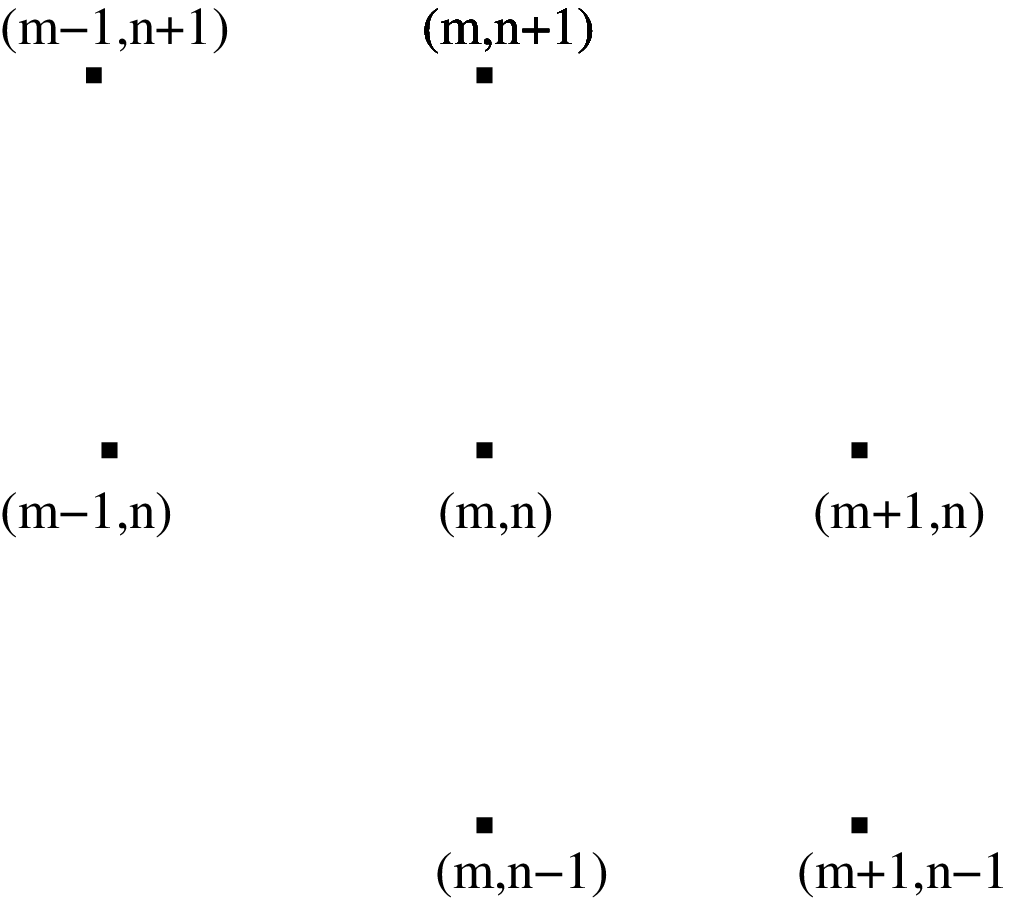}}}
\end{center}

\noindent
Fig.1~~~~~~~~~~ The 5 - and 7 - point schemes on the square lattice  

\vskip 10pt

\section{Basic criteria for discretizing partial differential operators}
\label{Criteria}

In order to construct the proper discretization of a partial 
differential operator we are guided by the following criteria. 

\begin{enumerate}

\item It should possess a large class of (discrete, continuous, isospectral, 
nonisospectral,..) symmetries, (at least) as large as that of its differential  
counterpart.

\item Its spectral properties should be similar to those of its differential 
counterpart.

\item The discretization should preserve the hyperbolic or elliptic 
character of the partial differential operator; in particular, if the operator 
is elliptic, the discretization should be applicable to solve a generic Dirichlet 
boundary value problem on a 2D lattice.

\item If the continuous operator is geometrically significant, the 
discretization should possess a geometric meaning which generalizes 
naturally that of the continuous operator.

\end{enumerate} 

In this paper we show that the difference operators \mref{L7}, 
\mref{L5} are discretizations, respectively, of the partial differential operators 
\mref{selffc}, \mref{selfc} 
which satisfy the properties 1., 3. and 4. (the spectral properties will be 
discussed elsewhere).

\section{The Dirichlet boundary value problem}
\label{BV}
As we pointed out in Section \ref{Criteria}, 
a proper discretization of a second order  elliptic operator 
should be applicable to solve Dirichlet boundary value problems on a 2D lattice. 

Consider, for the sake of concreteness, the following Dirichlet 
problem on a bounded domain of ${\mathbb R}^2$ for the operator (\ref{selfc}): 
\begin{equation}
\label{selfadj}
\begin{array}{l}
(A \Psi,_x),_x+(B \Psi,_y),_y =
F\Psi,~~~(x,y)\in{\cal D}\subset {\mathbb R}^2, \\
\Psi(x,y) ~~\hbox{given on}~~ \partial{\cal D},
\end{array}
\end{equation}
which appears very frequently in applications. 

It is easy to convince one-self that the 5-point self-adjoint scheme 
for the operator $\mathcal{L}_{5}$ (see e.q. (\ref{L5})): 
\begin{equation}
\label{saeq}
a_{m,n} \psi_{m+1,n} +a_{m-1,n}\psi_{m-1,n}
+b_{m,n} \psi_{m,n+1} +b_{m,n-1}\psi_{m,n-1}=
f_{m,n} \psi_{m,n} 
\end{equation}
which, in the natural continuous limit, reduces to the above equation (\ref{selfadj}), 
is perfectly adequate to solve a generic Dirichlet 
boundary value problems on a 2D lattice. The reasoning behind it   
(well-known to numerical analysts \cite{Hil})  
is clarified by the illustrative example of Figure 2. 
\begin{center}
\mbox{\hsize2cm \vbox{\epsfxsize=3cm 
\epsffile{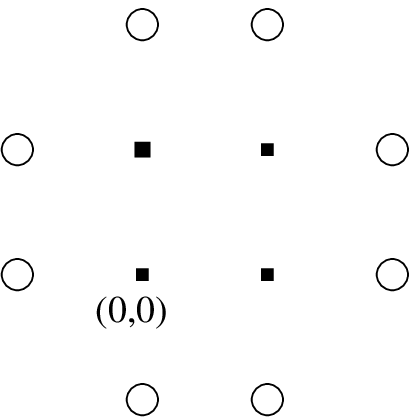}}}
\end{center}
Fig.2 $~~~~~~~~$ A simple Dirichlet problem for the 5-point scheme
\vskip 5pt
Suppose we want to solve the Dirichlet problem associated with the 5-point scheme 
(\ref{saeq}) in the subset of ${\mathbb Z}^2$ consisting of the white and black points in Fig.2. 
If the field $\psi_{m.n}$ is given at the boundary points (the 
white points), the unknown values of $\psi_{m.n}$ at the 4 interior points (the black points) 
are {\bf uniquely} constructed solving the following linear, inhomogeneous, determined system 
of 4 equations for 4 unknowns:         
\begin{equation}
\begin{array}{l}
\label{Lpsi}
-f_{0,0}\psi_{0,0}+a_{0,0}\psi_{1,0}+b_{0,0}\psi_{0,1}=-a_{-1,0}\psi_{-1,0}-b_{0,-1}\psi_{0,-1} \\
a_{0,0}\psi_{0,0}-f_{1,0}\psi_{1,0}+b_{1,0}\psi_{1,1}=-a_{1,0}\psi_{2,0}-b_{1,-1}\psi_{1,-1}  \\
b_{0,0}\psi_{0,0}-f_{0,0}\psi_{0,1}+a_{0,1}\psi_{1,1}=-a_{-1,1}\psi_{-1,1}-b_{0,1}\psi_{0,2}  \\
b_{1,0}\psi_{1,0}+a_{0,1}\psi_{0,1}-f_{1,1}\psi_{1,1}=-a_{1,1}\psi_{2,1}-b_{1,1}\psi_{1,2}
\end{array} 
\end{equation}
obtained applying 4 times the 5-point scheme (\ref{saeq}) with center at the interior 
points.

The same argument holds for more general subsets of ${\mathbb Z}^2$;   
its only possible failure is associated with the non generic situation in which the relevant 
matrix determinant of the system (which depends on the coefficients $a,b,f$) is zero.  


The definitions of interior and boundary points used in the above illustrative 
example are intuitive: the (nearest) neighbourhood of a point $(m,n)$ of the 
square lattice consists of the four points $(m+1,n)$, $(m,n+1)$, $(m-1,n)$, $(m,n-1)$. 
Given a subset $\Omega$ of ${\mathbb Z}^2$, its interior points are the 
points of $\Omega$ for which all neighbouring points belong to $\Omega$; its 
boundary points $\partial \Omega$ are instead the points of $\Omega$ such that 
some of the  neighbouring points 
do not belong to $\Omega$.

We remark that the 5-point scheme (\ref{saeq}) is, 
among all possible difference equations adequate to solve 
Dirichlet problems on 2D lattices, the {\bf simplest} possible scheme.

Using similar considerations, one can show that the 7-point scheme
\begin{equation}
\label{saeq2}
\begin{array}{l}
a_{m,n} \psi_{m+1,n} +a_{m-1,n} \psi_{m-1,n}
+b_{m,n} \psi_{m,n+1} +b_{m,n-1}  \psi_{m,n-1}+\\
s_{m+1,n} \psi_{m+1,n-1}+s_{m,n+1} \psi_{m-1,n+1}=
f_{m,n} \psi_{m,n},
\end{array}
\end{equation}
is applicable to solve Dirichlet problems on a 2D lattice.
Notice that, on a square lattice, two white points  should be added to the boundary with 
respect to 5-point scheme.
\begin{center}
\mbox{\hsize2cm \vbox{\epsfxsize=3cm 
\epsffile{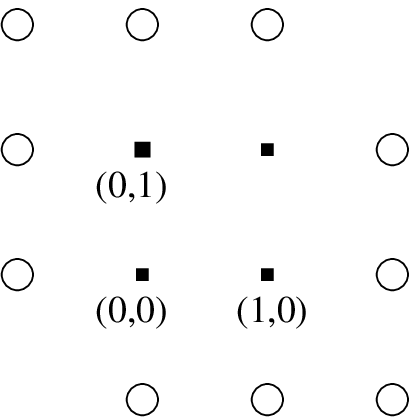}}}
\end{center}
\begin{center}
Fig.3 ~~   A Dirichlet problem for the 7-point scheme on the square lattice
\end{center}
\vskip 5pt

\section{Darboux transformations}
\label{SAD}

Non isospectral symmetries of Darboux type for linear differential 
operators play a relevant role in the theory of nonlinear integrable systems. 
They allow, for instance, to construct solutions of these nonlinear systems 
from simpler solutions, through an iterative procedure. As we mentioned in Section 
\ref{Criteria}, a good discretization of a partial differential operator should 
preserve this type of symmetries.    

In this section we present the DTs for the operators ${\mathcal L}_{5}$ and ${\mathcal L}_{7}$. 
These results are extracted from \cite{NDS}. 

\subsection{DTs for ${\mathcal L}_{5}$}
Consider the operator ${\mathcal L}_{5}$ 
together with the associated difference equation: 
\begin{equation}
\label{saeq-2}
a_{m,n} \psi_{m+1,n} +a_{m-1,n} \psi_{m-1,n}
+b_{m,n} \psi_{m,n+1} +b_{m,n-1}  \psi_{m,n-1}=f_{m,n} \psi_{m,n},
\end{equation}
where  $a_{m,n}$,  $b_{m,n}$ and $f_{m,n}$ are  given functions.

The operator ${\mathcal L}_{5}$ exhibits the following covariance property 
(gauge invariance): 
\begin{equation}
\begin{array}{c}
\label{gaugea}
{\mathcal L}_{5} \to \tilde{\mathcal L}_{5} =  
g_{m,n} {\mathcal L}_{5} g_{m,n} \\
a_{m,n} \to \tilde{a}_{m,n} = a_{m,n} g_{m,n} g_{m+1,n}, \qquad
b_{m,n} \to \tilde{b}_{m,n} = b_{m,n} g_{m,n} g_{m,n+1}, \\
f_{m,n} \to \tilde{f}_{m,n} = f_{m,n} g^2_{m,n}
\end{array}
\end{equation}
and possesses the following DTs.

\vskip 5pt
Let  $\tH$ be another solution of (\ref{saeq-2}), i.e:
\begin{equation}
\label{normal1}
a_{m,n} \tH_{m+1,n} +a_{m-1,n} \tH_{m-1,n}
+b_{m,n} \tH_{m,n+1} +b_{m,n-1}  \tH_{m,n-1}=f_{m,n} \tH_{m,n};
\end{equation}
then:
\begin{equation}
\label{ff}
f_{m,n}=\frac{1}{\tH_{m,n}}\left(a_{m,n} \tH_{m+1,n} +a_{m-1,n} \tH_{m-1,n}
+b_{m,n} \tH_{m,n+1} +b_{m,n-1}  \tH_{m,n-1}\right).
\end{equation}
Eliminating $f_{m,n}$ from \mref{saeq-2} and  \mref{normal1} we get
\begin{equation}
\begin{array}{l}
\Di (a_{m-1,n} \psi_{m,n}  \tH_{m-1,n} -a_{m-1,n} \tH_{m,n}  \psi_{m-1,n})+\\
\Dj (b_{m,n-1} \psi_{m,n}  \tH_{m,n-1}- b_{m,n-1} \tH_{m,n}  \psi_{m,n-1})=0,
\end{array}
\end{equation}
where
\begin{displaymath}
\begin{array}{l}
\Di f_{m,n}:=f_{m+1,n}-f_{m,n} \qquad \Dj f_{m,n}:=f_{m,n+1}-f_{m,n} \\
 \Dim f_{m,n}:=f_{m-1,n}-f_{m,n} \qquad \Djm f_{m,n}:=f_{m,n-1}-f_{m,n}.
\end{array}
\end{displaymath}
It means that there exists  a function $\alpha$ such that
\begin{equation}
\begin{array}{l}
\Dj \alpha = a_{m-1,n} \tH_{m,n} \tH_{m-1,n} 
\Dim \frac{\psi_{m,n}}{\tH_{m,n}},
\\
\\
\Di \alpha = -b_{m,n-1} \tH_{m,n}  \tH_{m,n-1} 
\Djm \frac{\psi_{m,n}}{\tH_{m,n}}, 
\end{array}
\end{equation}
where:
\begin{displaymath}
\begin{array}{l}
\Di f_{m,n}=f_{m+1,n}-f_{m,n} \qquad \Dj f_{m,n}=f_{m,n+1}-f_{m,n} \\
 \Dim f_{m,n}=f_{m-1,n}-f_{m,n} \qquad \Djm f_{m,n}=f_{m,n-1}-f_{m,n}.
\end{array}
\end{displaymath}

\noindent
Setting \[\psi'_{m,n}=\frac{\alpha_{m,n}}{\tH_{m,n}},\]
we find that $\psi'_{m,n}$ satisfies the following equation
\begin{equation}
\label{saeqp}
 a'_{m,n} \,  \psi'_{m+1,n} + a'_{m-1,n} \,  
\psi'_{m-1,n} +  b'_{m,n} \,  \psi'_{m,n+1}+ b'_{m,n-1} \, 
 \psi'_{m,n-1} =f'_{m,n} \, \psi'_{m,n}, 
\end{equation}
where
\begin{equation}
a'_{m-1,n}=\frac{\tH_{m,n}}{ b_{m-1,n-1} \,   \tH_{m-1,n-1}} \qquad
b'_{m,n-1}=\frac{\tH_{m,n}}{ a_{m-1,n-1} \,  \tH_{m-1,n-1}}
\end{equation}
and
\begin{equation}
\label{ff'}
f'_{m,n}=\tH_{m,n} \left( 
 a'_{m,n} \,  \frac{1}{\tH_{m+1,n}} + a'_{m-1,n} \,  \frac{1}{\tH_{m-1,n}} +
  b'_{m,n} \,  \frac{1}{\tH_{m,n+1}}+ b' _{m,n-1} \,  
\frac{1}{\tH_{m,n-1}}\right).
\end{equation}
Comparing equations (\ref{ff}) and (\ref{ff'}), we also infer that 
$\theta'=1/\theta$ is a solution of (\ref{saeqp}).

\subsection{DTs for ${\mathcal L}_{7}$}
\label{SAD7}
The construction of DTs presented in 
the previous  sub-section applies to the  
self-adjoint 7-point scheme associated with ${\mathcal L}_{7}$:
\begin{equation}
\label{sa7-2}
\begin{array}{l}
a_{m,n} \psi_{m+1,n} +a_{m-1,n} \psi_{m-1,n}
+b_{m,n} \psi_{m,n+1} +b_{m,n-1}  \psi_{m,n-1}+\\
s_{m+1,n} \psi_{m+1,n-1}+s_{m,n+1} \psi_{m-1,n+1}=
f_{m,n} \psi_{m,n},
\end{array}
\end{equation}
which is a discretization of the most general second order, 
self-adjoint, linear, differential equation in two independent variables.

Let  $\tH_{m,n}$ be another solution of equation \mref{sa7-2}:
\begin{equation}
\label{sas7}
\begin{array}{l}
a_{m,n} \tH_{m+1,n} +a_{m-1,n} \tH_{m-1,n}
+b_{m,n} \tH_{m,n+1} +b_{m,n-1}  \tH_{m,n-1}
+\\
s_{m+1,n} \tH_{m+1,n-1}+s_{m,n+1} \tH_{m-1,n+1}
=f_{m,n} \tH_{m,n}.
\end{array}
\end{equation}
Eliminating $f_{m,n}$ from \mref{sa7-2} and  \mref{sas7} we get
\begin{equation}
\begin{array}{l}
\Di [a_{m-1,n}  \tH_{m,n}  \tH_{m-1,n} (\frac{\psi_{m,n}}{\tH_{m,n}} - 
\frac{\psi_{m-1,n}}{\tH_{m-1,n} })
+ s_{m,n} \tH_{m-1,n} \tH_{m,n-1} (\frac{\psi_{m,n-1}}{\tH_{m,n-1}} - 
\frac{\psi_{m-1,n}}{\tH_{m-1,n} })]+
\\
\\
\Dj [b_{m,n-1} \tH_{m,n}   \tH_{m,n-1} (\frac{\psi_{m,n}}{\tH_{m,n}}- 
\frac{\psi_{m,n-1}}{\tH_{m,n-1}})
+s_{m,n} \tH_{m-1,n} \tH_{m,n-1} (\frac{\psi_{m-1,n}}{\tH_{m-1,n}}- 
\frac{\psi_{m,n-1}}{\tH_{m,n-1}})]=0.
\end{array}
\end{equation}
It means that there exists  a function $\psi'$ such that
\begin{equation}
\begin{array}{l}
\Dj ( \psi'_{m,n} \tH_{m,n})= \left( a_{m-1,n} \tH_{m,n} \tH_{m-1,n} + s_{m,n} 
\tH_{m-1,n} \tH_{m,n-1}   \right) 
\Dim \frac{\psi_{m,n}}{\tH_{m,n}}- \\
s_{m,n} \tH_{m-1,n} \tH_{m,n-1} 
\Djm \frac{\psi_{m,n}}{\tH_{m,n}}
\\
\\
\Di (\psi'_{m,n} \tH_{m,n}) = -(b_{m,n-1} \tH_{m,n}  \tH_{m,n-1} +
s_{m,n} \tH_{m-1,n} \tH_{m,n-1} )
\Djm \frac{\psi_{m,n}}{\tH_{m,n}}+ \\
s_{m,n} \tH_{m-1,n} \tH_{m,n-1} 
\Dim \frac{\psi_{m,n}}{\tH_{m,n}}.
\end{array}
\end{equation}
and function $\psi'_{m,n}$ satisfies the following equation
\begin{equation}
\begin{array}{l}
\label{bb}
 a'_{m,n} \,  \psi'_{m+1,n} + a'_{m-1,n} \,  \psi'_{m-1,n} +  b'_{m,n} \,  \psi'_{m,n+1}+
 b'_{m,n-1} \, 
 \psi'_{m,n-1} +\\
s'_{m+1,n} \psi'_{m+1,n-1}+s'_{m,n+1} \psi'_{m-1,n+1}
=f'_{m,n} \, \psi'_{m,n}, 
\end{array}
\end{equation}
where the new fields are given by
\begin{equation}
\begin{array}{l}
a'_{m,n}=\frac{\tH_{m,n} \tH_{m+1,n} a_{m-1,n}}{\tH_{m,n-1}  p_{m,n}},  
\qquad
b'_{m,n}=\frac{\tH_{m,n} \tH_{m,n+1} b_{m,n-1}}{\tH_{m-1,n}  p_{m,n}},
\\
\\
s'_{m,n}=\frac{s_{m-1,n-1} \tH_{m-1,n} \tH_{m,n-1} }
{\tH_{m-1,n-1} p_{m-1,n-1}}, 
\\
\\
f'_{m,n}=\tH_{m,n} (a'_{m,n} \frac{1}{\tH_{m+1,n}}+   
a'_{m-1,n} \frac{1}{\tH_{m-1,n}}+  
b'_{m,n} \frac{1}{\tH_{m,n+1}} + b'_{m,n-1} \frac{1}{\tH_{m,n-1}}+
\\
\\
s'_{m+1,n} \frac{1}{\tH_{m+1,n-1}}+s'_{m,n+1} \frac{1}{\tH_{m-1,n+1}})
\end{array}
\end{equation}
and where
$p_{m,n}= \tH_{m,n}a_{m-1,n}b_{m,n-1} +
\tH_{m-1,n} s_{m,n} a_{m-1,n}+s_{m,n} \tH_{m,n-1} b_{m,n-1}$. Again 
$\tH'_{m,n}=1/\tH_{m,n}$ is a solution of \mref{bb}.

\section{Lelieuvre formulas}
\label{CL}
Lelieuvre idea of describing the surface  parametrized with
asymptotic co-ordinates via its co-normal
image \cite{Lelieuvre,Blaschke} plays an important role in the
theory of  surfaces in equi-affine  spaces.
Due to the generalizations of Lelieuvre formulas
to a co-ordinate free language  \cite{LiNo,Li},  
one can describe the hyper-surface via its co-normal image
in an arbitrary coordinate system.

In this section we show that the co-normal image of a general 2D 
surface in $eA^3$ is a vector solution of the partial differential 
equation
\begin{equation}
\label{eq7c}
(A \Psi,_x)_x+(S \Psi,_y),_x+ (B \Psi,_y),_y +(S \Psi,_x),_y=F \Psi
\end{equation}
associated with the operator (\ref{selffc}).

We recall some basic facts.
We denote  by $\R$ the position
vector $\R : {\mathbb R^2} \rightarrow eA^3$ of a parametrized surface in $eA^3$ 
and we assume that the surface 

\noindent
i) be regular, i.e.: 
\begin{equation}
\label{c1}
\N \propto \R,_x \times \R,_y \ne 0; 
\end{equation}
ii) be twice differentiable; i.e., in particular:
\begin{equation}
\label{c2}
\R,_{xy}=\R,_{yx};
\end{equation}
iii) be locally strongly convex, hence:
\begin{equation}
\label{c3} 
Vol^*(\N ; \N,_x ; \N,_y) \ne 0.
\end{equation}
Then, from \mref{c3}, we infer that
the vector fields $\N \times \N,_x$ and $\N \times \N,_y$
are linearly independent and are tangent field to the
surface.
Therefore one can decompose the fields $\R,_x$ and $\R,_y$
as follows:
\begin{equation}
\begin{array}{c}
\R,_y= A \N \times \N,_x + P \N \times \N,_y, \\
\R,_x= -Q \N \times \N,_x -B \N \times \N,_y,
\end{array}
\end{equation}
where,  
 since $\R,_x \times \R,_y=(AB-PQ) Vol^*(\N ; \N,_x ; \N,_y) \N$,
due to assumption \mref{c1}, 
we have $AB-PQ \ne 0$.
The equality $<\N|\R,_{xy}-\R,_{yx}>=0$ gives
$(P-Q) Vol^*(\N ; \N,_x ; \N,_y)=0 $; so we have $P=Q=:S$
and, finally, 
\begin{equation}
\label{L}
\begin{array}{c}
\R,_y= A \N \times \N,_x + S \N \times \N,_y, \\
\R,_x= -S \N \times \N,_x -B \N \times \N,_y,
\end{array}
\end{equation}
\begin{equation}
AB-S^2 \ne 0.
\end{equation}
The compatibility condition $\R,_{xy}=\R,_{yx}$ of equations \mref{L} leads 
to the partial differential equation
\begin{equation}
\label{selfadv}
(A \N,_x)_x+(S \N,_y),_x+ (B \N,_y),_y +(S \N,_x),_y=F \N
\end{equation}
associated with the operator (\ref{selffc}), 
which is nothing but the most general self-adjoint equation
of second order in two independent variables.

Conversely, let $N_1$, $N_2$ and $N_3$ be three linearly independent solutions 
of the self-adjoint equation \mref{eq7c}, which we assume to be non parabolic; i.e.: 
\begin{equation}
AB-S^2 \ne 0.
\end{equation} 
Select any frame in $eA^3$ and the vector field
$\N=[N_1,N_2,N_3]$ with respect to the co-frame. Since $N_1$, $N_2$
and $N_3$  are linearly independent, we have that $Vol^*(\N;\N,_x;\N,_y) \ne 0$. 
In addition $\N$ satisfies equation \mref{selfadv}; 
the vector multiplication of both sides of this equation 
for $\N$ by $\N$ it-self yields, after manipulation, the equation
\[(A \N \times \N,_x + S \N \times \N,_y),_x+
(S \N \times \N,_x + B \N \times \N,_y),_y=0 ,\]
from which we infer that there exists a vector field $\R$
such that equations \mref{L} hold.
Interpreting $\R$ as the position vector of
a surface, we infer that $\N$ is a co-normal field to this 
surface and that this surface is regular, since
\[ \R,_x \times \R,_y=(AB-S^2) Vol^*(\N ; \N,_x ; \N,_y) \N . \]

\section{Lelieuvre formulas associated with the 7-point scheme} 
\label{GL}

In the previous section we have shown that the operator (\ref{selffc}) characterizes the co-normal 
image of a generic surface in $eA^3$. According to the last criterion 
of Section \ref{Criteria}, a good discretization of (\ref{selffc})  should possess an analogous 
geometric meaning. Indeed in this section we will show that the difference 
operator ${\mathcal L}_7$ describes the co-normal image of a generic 2D lattice in $eA^3$.

Consider a lattice ${\mathbb Z^2} \supset \Omega \cup \partial \Omega \rightarrow eA^3$ and  
denote by $\rr_{m,n}$ the position vector with respect to a frame.
By "lower" triangles we mean the triangles with
vertices $(\rr_{m,n},\rr_{m+1,n},\rr_{m,n+1})$ and 
by "upper" triangles we mean the triangles
with vertices $(\rr_{m+1,n+1},\rr_{m+1,n},\rr_{m,n+1})$.

\begin{center}
\mbox{\vsize7cm \vbox{\epsfxsize=10cm 
\epsffile{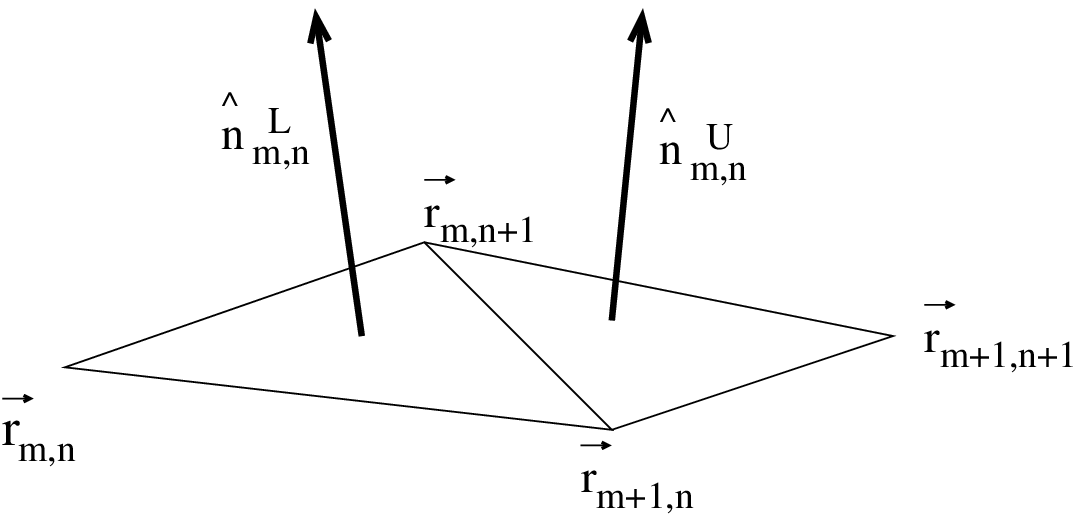}}}
\end{center}

\noindent 
Fig.4 ~~~~~~~~~~ Upper and lower triangles of the 2D lattice

\vskip 5pt

We assume that:
\begin{itemize}
\item 
The upper and lower triangles are not degenerate; i.e., the 
three points of each triangle are not collinear. For the 
lower triangles of the 2D lattice this condition means that:
\begin{equation}
\label{regl}
\Di \rr_{m,n} \times \Dj \rr_{m,n} \ne 0.
\end{equation}
Then we denote by $\nn^L_{m,n}$ any co-normal non-vanishing field to the lower triangles:
\[\nn^L_{m,n}: =\lambda^L_{m,n}  \Di \rr_{m,n} \times \Dj \rr_{m,n},\] where
$\lambda^L_{m,n}$ is a non-vanishing scalar field. Analogously, the non degeneracy condition 
\begin{equation}
\label{regu}
\Di \rr_{m,n+1} \times \Dj \rr_{m+1,n} \ne 0
\end{equation}
for the upper triangles of the 2D lattice allows one to define  
any co-normal non-vanishing field $\nn^U_{m,n}$ to the upper triangles by:
\[\nn^U_{m,n}:= \lambda^U_{m,n} \Di \rr_{m,n+1} \times \Dj \rr_{m+1,n} , \] 
where $\lambda^U_{m,n}$ is a non-vanishing scalar field.
\item
The fields $\nn^L_{m,n}$ and $\nn^U_{m,n}$
satisfy the following conditions:
\end{itemize}
\begin{equation}
\label{conl}
V^L_{m,n}:=Vol^*(\nn^L_{m,n}, \nn^L_{m+1,n},\nn^L_{m,n+1}) \ne 0 ,
\end{equation}
\begin{equation}
\label{conu}
V^U_{m,n}:=Vol^*(\nn^U_{m,n}, \nn^U_{m-1,n},\nn^U_{m,n-1}) \ne 0.
\end{equation}

>From assumption \mref{conl} it follows that 
the  discrete vector fields 
$\nn^L_{m,n} \times \nn^L_{m-1,n}$ and  $\nn^L_{m,n} \times \nn^L_{m-1,n+1}$ 
 are linearly
independent and therefore they span the tangent space to the
lower triangle 
(the same is true for fields
$\nn^L_{m,n} \times \nn^L_{m,n-1}$ and  $\nn^L_{m,n} \times
\nn^L_{m+1,n-1}$).
So we can write
\begin{equation}
\begin{array}{l}
\Dj \rr_{m,n}=\nn^L_{m,n} \times 
(a_{m-1,n} \nn^L_{m-1,n}+p_{m-1,n} \nn^L_{m-1,n+1}), \\
\Di \rr_{m,n}=-\nn^L_{m,n} \times 
(b_{m,n-1} \nn^L_{m,n-1}+q_{m,n-1} \nn^L_{m+1,n-1}).
\end{array}
\end{equation}
The equality 
$<\nn^L_{m,n}| \Di \Dj \rr_{m,n}>=<\nn^L_{m,n}| \Dj \Di \rr_{m,n}>$
is equivalent to $ (p-q) V^L_{m,n}=0$; 
so we have (taking into account \mref{conl}) $p=q=:s$ and, as a result, 
\begin{equation}
\begin{array}{l}
\Dj \rr_{m,n}=\nn^L_{m,n} \times 
(a_{m-1,n} \nn^L_{m-1,n}+s_{m-1,n} \nn^L_{m-1,n+1}),\\
\Di \rr_{m,n}=-\nn^L_{m,n} \times 
(b_{m,n-1} \nn^L_{m,n-1}+s_{m,n-1} \nn^L_{m+1,n-1}),
\end{array}
\end{equation}
where the coefficients $a,b,s$ are defined by:
\begin{equation}
\label{abs}
\begin{array}{l}
a_{m,n}=-\frac{<\nn^L_{m,n+1}|\Delta_n\rr_{m+1,n}>}{V^L_{m,n}},~~
b_{m,n}=-\frac{<\nn^L_{m+1,n}|\Delta_n\rr_{m,n+1}>}{V^L_{m,n}}, \\
s_{m,n}=\frac{<\nn^L_{m,n}|\Delta_n\rr_{m+1,n}>}{V^L_{m,n}}=\frac{<\nn^L_{m,n}|\Delta_m\rr_{m,n+1}>}{V^L_{m,n}}.
\end{array}
\end{equation}
>From $\Di \Dj \rr_{m,n}=\Dj \Di \rr_{m,n}$ we finally get that the lower co-normal
vector satisfies the self-adjoint 7-point scheme
\begin{equation}
\label{se}
\begin{array}{l}
a_{m,n} \nn^L_{m+1,n}+a_{m-1,n} \nn^L_{m-1,n}+
b_{m,n}\nn^L_{m,n+1}+b_{m,n-1} \nn^L_{m,n-1}+\\
s_{m-1,n} \nn^L_{m-1,n+1}
+s_{m,n-1} \nn^L_{m+1,n-1}
=f_{m,n}\nn^L_{m,n}.
\end{array}
\end{equation}
Consider now the co-vector fields:
\begin{displaymath}
\begin{array}{l}
X_{m,n}:=a_{m,n} \nn^L_{m+1,n}+b_{m,n}\nn^L_{m,n+1}, \\
Y_{m,n}:=a_{m-1,n} \nn^L_{m-1,n}+s_{m-1,n} \nn^L_{m-1,n+1}, \\
Z_{m,n}:=b_{m,n-1} \nn^L_{m,n-1}+s_{m,n-1} \nn^L_{m+1,n-1};
\end{array}
\end{displaymath}
then equation \mref{se} can be re-written in terms of them:
\begin{equation}
X_{m,n}+Y_{m,n}+Z_{m,n}=f_{m,n}\nn^L_{m,n}.
\end{equation}
A direct calculation shows that
\begin{equation}
\Di \rr_{m,n} \times \Dj \rr_{m,n} =-V^L_{m,n} Vol^*
(\nn^L_{m,n};Y_{m,n};Z_{m,n}) \nn^L_{m,n} ,
\end{equation}
from which we infer that
\begin{equation}
\label{con3}
 Vol^*(\nn^L_{m,n};Y_{m,n};Z_{m,n}) \ne 0 .
\end{equation}
For the normal to the upper triangle:
\begin{equation}
\begin{array}{l}
\nn^U_{m,n} = \lambda^U_{m,n}
\Dj \rr_{m+1,n} \times \Di \rr_{m,n+1} =\\
 \lambda^U_{m,n}
V^L_{m,n} (a_{m,n} b_{m,n} \nn^L_{m,n}+a_{m,n} s_{m,n} \nn^L_{m+1,n}+b_{m,n} s_{m,n} \nn^L_{m,n+1}) ;
\end{array}
\end{equation}
so 
\begin{equation}
\label{cond1}
a_{m,n} b_{m,n} \ne 0 \qquad \hbox{or}  \qquad a_{m,n} s_{m,n} \ne 0 \qquad  \hbox{or} \qquad b_{m,n} s_{m,n} \ne 0 ,
\end{equation}
\begin{equation}
\begin{array}{l}
V^U_{m,n}= \lambda^U_{m,n} V^L_{m,n}
\lambda^U_{m-1,n} V^L_{m-1,n}\lambda^U_{m,n-1} V^L_{m,n-1}
Vol^*(\nn^L_{m,n};Y_{m,n};Z_{m,n}) *\\
(a_{m,n} a_{m-1,n} b_{m,n} b_{m,n-1}+
a_{m,n-1} a_{m-1,n} s_{m,n} b_{m-1,n}+ b_{m,n-1} b_{m-1,n} s_{m,n} s_{m,n-1}) .
\end{array}
\end{equation}
Therefore
\begin{equation}
\begin{array}{l}
\label{cond2}
a_{m,n} a_{m-1,n} b_{m,n} b_{m,n-1}+
a_{m,n-1} a_{m-1,n} s_{m,n} s_{m-1,n}+ b_{m,n-1} b_{m-1,n} s_{m,n} s_{m,n-1} \ne 0. 
\end{array}
\end{equation}

Summarizing,  we have the following theorem.
\begin{thm}
Consider a two-dimensional lattice ${\mathbb Z} \ni \Omega \rightarrow eA^3$ 
such that its position vector $\rr_{m,n}$ and its lower $\nn^L_{m,n}$ and upper $\nn^U_{m,n}$  co-normals 
obey the conditions \mref{regl}, \mref{regu}, \mref{conl} and \mref{conu}. 
Then there exist functions $a_{m,n}$, $b_{m,n}$ and  $s_{m,n}$ obeying conditions
\mref{cond1} and \mref{cond2} such that the following Lelieuvre type relations hold
\begin{equation}
\begin{array}{l}
\label{DLel}
\Dj \rr_{m,n}=\nn^L_{m,n} \times 
(a_{m-1,n} \nn^L_{m-1,n}+s_{m-1,n} \nn^L_{m-1,n+1}),\\
\Di \rr_{m,n}=-\nn^L_{m,n} \times 
(b_{m,n-1} \nn^L_{m,n-1}+s_{m,n-1} \nn^L_{m+1,n-1}),
\end{array}
\end{equation}
and such that the lower co-normal field satisfies the 7-point self-adjoint scheme

\begin{equation}
\label{DLNE}
\begin{array}{l}
a_{m,n} \nn^L_{m+1,n}+a_{m-1,n} \nn^L_{m-1,n}+
b_{m,n}\nn^L_{m,n+1}+b_{m,n-1} \nn^L_{m,n-1}+ \\
s_{m-1,n} \nn^L_{m-1,n+1}
+s_{m,n-1} \nn^L_{m+1,n-1}
=f_{m,n}\nn^L_{m,n}.
\end{array}
\end{equation}

Conversely, consider the field $\nn^L_{m,n}$ satisfying: i) equation \mref{DLNE} with the 
coefficients obeying conditions \mref{cond1} and \mref{cond2}; ii) 
the conditions \mref{conl} and \mref{con3}. Then the Lelieuvre type formulas 
\mref{DLel} define the position vector  $\rr_{m,n}$ of a 2D lattice in $eA^3$, having 
$\nn^L_{m,n}$ as a lower co-normal. The position vector, the lower co-normal and the upper
co-normal $\nn^U_{m,n}$ given by 
$\nn^U_{m,n}:= \lambda^U_{m,n} \Di \rr_{m,n+1} \times \Dj \rr_{m+1,n}$
satisfy the conditions \mref{regl}, \mref{regu} and \mref{conu}.
\end{thm}


\section{Lelieuvre formulas associated with the self-adjoint 5-point scheme} 
\label{geom}

In the previous two sections we have shown that, 
on the level of the Lelieuvre type description, 
the operator \mref{selffc} and its discretization ${\mathcal L}_7$ characterize 
respectively a generic 2D co-ordinate net and a generic 2D lattice in $eA^3$. 
In this section we introduce distinguished reductions on the above 
generic nets (lattices), showing that i) the reduction from a generic net to a 
conjugate net (a surface parametrized by conjugate co-ordinates) 
is characterized, on the level of the Lelieuvre type description, 
by the reduction from the general self-adjoint partial differential 
operator \mref{selffc} to the self-adjoint 
operator \mref{selfc}; ii) the reduction from a generic 2D lattice  to a 
2D quadrilateral lattice (a lattice whose elementary quadrilaterals are 
planar) is characterized, on the level of the Lelieuvre type description, 
by the reduction from the 7-point scheme \mref{saeq2} to the 5-point scheme \mref{saeq}.

Let $\R(x,y)$ be the position vector of a conjugate net and let 
$\N(x,y)$ be any co-normal vector field; then:
\[ <\N | \R_x> =0=<\N | \R_y>, ~~~~~~~\hbox{definition of $\N$}, 
\]
\[<\N |\R_{xy}>=0,
 ~~~~~~~~~~~~~~\hbox{conjugacy}. \]
Therefore, assuming that the surface be locally strongly convex (condition \mref{c3}), 
one infers that the following Lelieuvre type formulas hold:
\begin{equation}
\label{Lel3}
\R_x = B\N_y \times \N,~~~~\R_y = A\N \times \N_x ,
\end{equation}
where
\begin{equation}
B(x,y)=\frac{<\N_x | \R_x>}{Vol^*(\N,\N_x,\N_y)},~~
A(x,y)=\frac{<\N_y | \R_y>}{Vol^*(\N,\N_x,\N_y)}.
\end{equation}
The integrability condition $\R_{xy}=\R_{yx}$ implies equation 
\begin{equation}
(A\N_{xx}+A_x\N_x+B\N_{yy}+B_y\N_y)\times \N=0,
\end{equation}
which is equivalent to the wanted self-adjoint equation
\begin{equation}
\label{aa}
(A\N_{x})_x+(B\N_{y})_y=F\N,~~~~F=F(x,y).
\end{equation}
Conversely, it is straightforward to prove that, if $\N$ 
satisfies equation \mref{aa} for some coefficients $A,B,F$, 
then the vector $\R$, defined by \mref{Lel3}, 
is the position vector of a conjugate net having $\N$ as normal vector.

For the discrete case we follow the same reasoning. 
Any co-normal vector field of a quadrilateral lattice satisfies 
equations:
\[<\N_{m,n} | \Di \rr_{m,n}> =0=<\N_{m,n} | \Dj \rr_{m,n}>,~~~\hbox{definition of $\N$,}\]
\[<\N_{m,n} | \Di \Dj \rr_{m,n}> =0
,~~~~~~~~~~\hbox{quadrilaterality}.
\]

Therefore, assuming that 
\begin{equation}
Vol^*(\N_{m,n},\N_{m+1,n},\N_{m,n+1}) \neq 0,
\end{equation}
the following Lelieuvre type relations exist between the tangent vectors
and the co-normal to the lattice
\begin{equation}
\label{Lel5P}
\Di \rr_{m,n} = -b_{m,n-1} \N_{m,n} \times \N_{m,n-1},~~~~ 
\Dj \rr_{m,n} = a_{m-1,n} \N_{m,n} \times \N_{m-1,n},
\qquad 
\end{equation}
where the scalar fields $a_{m,n}$ and $b_{m,n}$ are defined by:
\begin{equation}
\begin{array}{l}
a_{m,n}= -\frac{<\N_{m,n+1} | \Dj \rr_{m+1,n} >}{Vol^*(\N_{m,n},\N_{m+1,n},\N_{m,n+1})},
\\
b_{m,n}= \frac{<\N_{m+1,n} | \Di \rr_{m,n+1}>}{Vol^*(\N_{m,n},\N_{m+1,n},\N_{m,n+1})}.
\end{array}
\end{equation}

The integrability condition for equations \mref{Lel5P} implies the equation
\begin{equation}
\label{normal}
(a_{m,n} \N_{m+1,n} +a_{m-1,n} \N_{m-1,n}
+b_{m,n} \N_{m,n+1 } +b_{m,n-1}  \N_{m,n-1}) \times  \N_{m,n}=0,
\end{equation}
which is equivalent to equation ${\mathcal L}_{5}\N=0$. 

Conversely, it is also easy to show
that, if  $\N_{m,n}$ satisfies equation \mref{normal}, then the Lelieuvre formulas
\mref{Lel5P} define a proper embedding of a 2D quadrilateral lattice having $\N_{m,n}$ as co-normal.

\vskip 5pt
\begin{center}
\mbox{
\vbox{\epsfxsize=10cm \epsffile{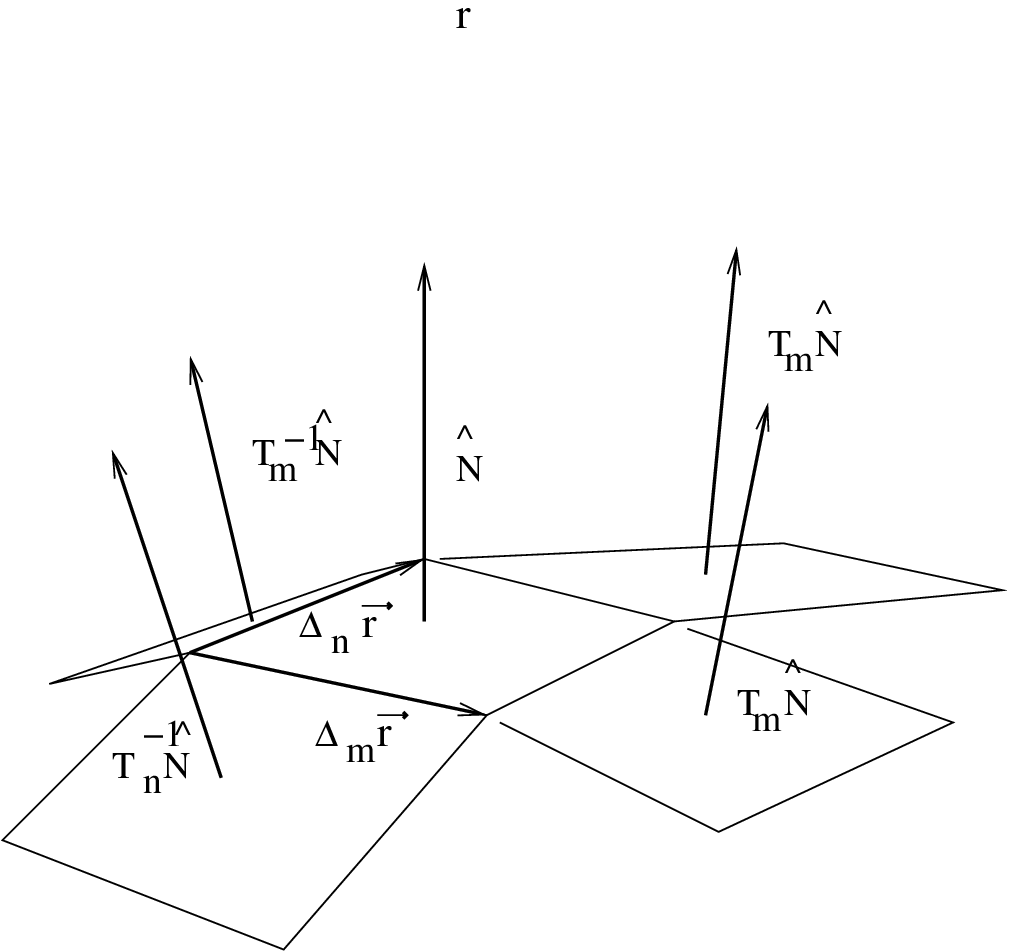}}
}
\end{center}

\noindent 
Fig.5 ~~~~~~~~~~~~~~~~~~The 5-point scheme for the normal vector

\vskip 5pt

We remark that this result could have been deduced in a faster way, f.i., from (\ref{abs}c), 
observing that the reduction from ${\mathcal L}_{7}$ to ${\mathcal L}_{5}$, expressed by the  
equation $s_{m,n}=0$, is equivalent to the condition that the tangent vectors of the 
upper triangles of the 2D lattice are perpendicular to $\hat n^L_{m,n}$; i.e., it is 
equivalent to the condition that the 2D lattice is quadrilateral.

We conclude this section remarking that, due to the gauge covariance properties of the 
operators \mref{selfc} and ${\mathcal L}_{5}$ (see \mref{gaugea} and its continuous limit), 
the normal vectors appearing in the characterizing equations \mref{aa} and ${\mathcal L}_{5}\N=0$ have an arbitrary 
normalization. The characterizations (see, e.g., \cite{DS}) 
\[\N_{xy}+\alpha\N_x+\beta\N_y=0, \]
\[\Di\Dj\N_{m,n}+\alpha\Di\N_{m,n}+\beta\Dj\N_{m,n}=0\]
of respectively a 2D conjugate net and of a 2D quadrilateral lattice 
in terms of their co-normal vectors are instead gauge dependent. 

\vskip 20pt
\noindent
{\bf Acknowledgments}
\vskip 10pt
\noindent
This work was supported by the cultural and scientific agreement between 
the University of Roma ``La Sapienza'' and the University of Warsaw
and partially supported by KBN grant 2 P03B 126 22.

\end{document}